# Impact of Fourth Industrial Revolution (4IR) on Small and Medium Enterprises (SMEs) and Employment in Bangladesh: Opportunities and Challenges

Toukir Ahammed
Institute of Information Technology (IIT)
University of Dhaka
Dhaka, Bangladesh
Email: toukir@iit.du.ac.bd

Moumita Asad
Institute of Information Technology (IIT)
University of Dhaka
Dhaka, Bangladesh
Email: moumita@iit.du.ac.bd

Kazi Sakib
Institute of Information Technology (IIT)
University of Dhaka
Dhaka, Bangladesh
Email: sakib@iit.du.ac.bd

## Abstract

*The Fourth Industrial Revolution (4IR) is transforming industries and economies worldwide, presenting both opportunities and challenges for Small and Medium Enterprises (SMEs) and employment. This study qualitatively explores the impact of 4IR on the SME sector in Bangladesh highlighting the associated opportunities and challenges. Initially, secondary data sources, including newspaper articles, research papers and government reports, are reviewed to establish the context of the study and to prepare the questionnaire for primary data collection. Then, the primary data is collected through Key Informant Interviews and Focus Group Discussions with different stakeholders including SME owners, association representatives, and government officials. The study reveals that while most of the participants have only a superficially awareness of 4IR, they view it positively as a blessing for the SME sector. Despite being in early adoption stages in Bangladesh, SMEs anticipate numerous benefits such as enhanced customer experiences, reduced production times, improved product quality, and data-driven decision-making. Regarding employment, most participants believe that adopting 4IR technologies in the SME sector of Bangladesh will create new job opportunities and boost overall productivity. However, participants express concern about challenges during the transition to 4IR in SME sector, including a lack of technical knowledge, financial constraints, inadequate training, safety and security issues, infrastructure limitations, and difficulties in accessing essential information. To fully harness 4IR's potential benefits for SMEs in Bangladesh, several key recommendations emerge that include analyzing of the current SME landscape comprehensively, establishing a collaborative information sharing platform, organizing effective training and workshops, promoting resource sharing, encouraging local innovation, attracting foreign clients, launching awareness campaigns, ensuring proper policy implementation and fostering collaboration among government entities, associations, and academia. By addressing these challenges and implementing the recommended strategies, Bangladesh can effectively embrace the transformative benefits of 4IR, simultaneously improving its SME sector and fostering positive employment outcomes.*

## 1 Introduction

A Small and Medium-sized Enterprise (SME) is a business entity that falls within a specific size range in terms of factors such as the number of employees, annual revenue, or total assets. The exact criteria defining an SME can vary by country and industry (Madani, 2018). In Bangladesh, enterprises are classified into Small and Medium Enterprises (SMEs) based on size and assets by the Ministry of Industries. In the manufacturing sector, small enterprises encompass assets valued from BDT 75 lakh to 15 crores, and/or 26 to 120 workers, while medium enterprises possess assets valued from BDT 15 crore to 50 crore (excluding land and factory building) and employ 121 to 300 workers. In the service sector, small enterprises have

16 to 50 employees and assets worth BDT 10 lakh to 2 crores, whereas medium enterprises employ 51 to 120 individuals and hold assets worth BDT 2 crore to 30 crores[1].

SMEs play a crucial and significant role in the economic development of Bangladesh (Al Asheq et al., 2021) and are often described as the backbone of the Bangladeshi economy (Atikur Rahaman & Sayed Uddin, 2022). These enterprises typically have a limited number of employees, resources and assets, but they are regarded as the main drivers of the economy for a variety of reasons such as the creation of jobs, increased production, the reduction of poverty, etc. Moreover, SMEs span a diverse range of industries, from startups to traditional family-owned businesses. According to the SME Foundation[2], there are 7.8 million SMEs (including Cottage and Micro) in Bangladesh which employ more than 21 million people and contribute to around 25% of the GDP of Bangladesh (Khan et al., 2021).

Given their significant economic role, it is necessary to consider how SMEs in Bangladesh can adapt to and benefit from the rapid technological changes brought about by the Fourth Industrial Revolution (4IR), also known as Industry 4.0. The 4IR refers to the rapid technological transformation of industries, societal patterns, processes and practices with the latest smart technology. Every industrial revolution has marked a significant shift in how society operates. In the 18$^{th}$ century, the invention of the steam engine led to the First Industrial Revolution by enabling mechanized production for the first time and resulting in a significant shift toward urbanization. During the Second Industrial Revolution, also known as the Technological Revolution, the widespread use of electricity and other scientific advancements led to the era of mass production. The Third Industrial Revolution, also known as the Digital Revolution, started in the late 20$^{th}$ century, brought about the rise of computers, digital technologies, and the internet resulting in the growing automation in manufacturing. Building on the foundation laid by three industrial revolutions, the fourth industrial revolution focuses on the integration of physical, digital, and biological systems.

4IR has changed the way we live, the way we think and the way we work. It has brought about radical transformations and disrupted almost every business sector including SMEs. It influences the way SMEs operate, innovate, and compete in today's rapidly evolving business requirements. The ongoing transformative phase is fueled by rapid technological advancements in areas like Artificial Intelligence (AI), automation, 3D printing, the Internet of Things (IoT), Cloud Computing, Blockchain, autonomous robots, Virtual Reality (VR), Augmented Reality (AR), biotechnology, etc. In the current situation, businesses need to meet Industry 4.0 requirements of delivering high-quality products at competitive prices in due time by adopting these advanced technologies.

While 4IR offers promising opportunities to the SMEs in Bangladesh to enhance their competitiveness through digitalization, automation and innovative business models, it also brings some inherent challenges in adoption such as a lack of skilled workforce, insufficient infrastructure due to limited access to finance and technology, etc. The adoption of 4IR technologies can free up human resources leaving the opportunities to employ these resources in other creative and strategic tasks.

Recent research about the fourth industrial revolution pays attention to the new and advanced technologies and how SMEs can adapt and utilize these technologies (Alqam & Saqib, 2020). The literature has focused on understanding how SMEs can adapt and utilize these technologies to remain competitive in today's rapidly changing environment. Some studies have identified the challenges that SMEs face in adopting and implementing 4IR technologies (Alqam &

---

[1] National Industry Policy of Bangladesh 2022
[2] http://smedata.smef.gov.bd/index.php/report/summaryTable

Saqib, 2020; Serumaga-Zake & van der Poll, 2021; Vuong & Mansori, 2021) whereas other studies highlight the opportunities presented by 4IR (Adegbite & Govender, 2021; Adelowotan, 2021; Alqam & Saqib, 2020). As SMEs are regarded as the backbone of the Bangladeshi economy, it is crucial for SMEs to perform up to the standard in the current economic situation of Bangladesh. Thus, it becomes essential to study the challenges of adopting 4IR technologies in the SME sector and embrace the opportunities offered by 4IR to ensure their contribution to the national economy of Bangladesh.

In this context, this paper aims to evaluate how 4IR will impact the SME sector and employment in Bangladesh in terms of opportunities and challenges. To fulfill this objective, the following research questions are addressed:

**RQ1. What is the meaning of 4IR in the context of SME?**

**RQ2. What technologies are driving 4IR in the SME sector?**

**RQ3. What opportunities are 4IR creating for the SMEs in Bangladesh?**

**RQ4. What challenges are the SMEs in Bangladesh facing due to 4IR?**

The first two questions aim to establish a foundational understanding of 4IR by asking about its meaning and the technologies that can drive the SME sector of Bangladesh. RQ3 and RQ4 will help to identify areas of the SME sector where policy interventions can be made to maximize the benefits of 4IR while mitigating its negative effects.

## 2 Literature Review

The emergence of the Fourth Industrial Revolution (4IR) has opened a new field of research in the context of Bangladesh, particularly when considering its impact on the country's Small and Medium-sized Enterprises (SMEs). In recent years, researchers have increasingly focused on understanding how 4IR technologies can be used to address local challenges and create opportunities for economic growth and social development in developing countries (Adegbite & Govender, 2021; Adelowotan, 2021; Alqam & Saqib, 2020; Serumaga-Zake & van der Poll, 2021; Vuong & Mansori, 2021).

The literature discussed in this section have been chosen for several reasons. Firstly, these studies provide an understanding of how 4IR technologies are being adopted and integrated into SMEs in developing countries. These studies also explore the unique challenges faced by SMEs in developing countries, such as limited access to technology, lack of skilled workforce, and financial constraints. Finally, these studies highlight the broader implications of 4IR for economic and social development, emphasizing the role of technology in driving innovation, improving productivity, and enhancing the quality of life. Additionally, a few studies have been reviewed to understand the current status and polices related to ICT adoption in the context of SMEs of Bangladesh.

Alqam & Saqib (2020) investigated the influence of the Fourth Industrial Revolution (4IR) on Small and Medium Enterprises (SMEs), particularly within the context of Oman. In this exploratory research, the authors collected primary data from the experts of top and middle management staff from several different SMEs through focus group discussions. The study also relied on previous literature related to 4IR and SME as the sources of secondary data. The results of the study suggest that all organizations need to initiate the development of an IT strategy to adopt 4IR concepts and technologies. The study pointed out some benefits that can enhance the competitive advantages of SMEs such as revenue, profitability, etc. by reducing the production time and meeting customer demand properly. However, the current technology infrastructure is one of the most important challenges for SMEs to adopt 4IR according to the study.

Serumaga-Zake & van der Poll (2021) identified the challenges faced by SMEs due to the business processes changed by 4IR. The focus of the study was only the manufacturing SMEs of South Africa. This study presents a comprehensive literature review focusing on small and medium enterprises (SMEs) and their significance in developing economies. It explores the unique needs and challenges faced by SMEs in contrast to larger, well-established organizations. The study also examines how SMEs can leverage new technologies despite their limited resources.

Adelowotan (2021) explored how SMEs can survive in the era of disruptive technologies of 4IR. The study examined the opportunities offered by the advanced technologies of 4IR in the context of SMEs in South Africa. The methodology of the study involves analyzing the 2018/2019 South Africa's Small and Medium Enterprises (SMEs) landscape survey results. This analysis aims to uncover the significant roles played by SMEs in the South African economy, pinpoint the challenges they face, and suggest how they can address these challenges through the adoption of 4IR technologies. Despite concerns about job loss and inequality due to 4IR technologies, the findings of the study suggest that with clear government policies and business community engagement, entrepreneurs can benefit from new supply chains and markets. Additionally, technological advancements will lead to increased productivity, consequently increasing labor demand which will create more new jobs than those displaced by automation.

A systematic literature review was carried out by Adegbite & Govender (2021) to investigate the extent of understanding concerning the adoption of 4IR technologies among SMEs, along with the benefits offered by these technologies to the SME sector in low-income nations, particularly in Africa. This study explores the potential roles of SMEs in the 4IR and its impact on sectoral growth and transformation in Africa by reviewing 29 relevant studies. The findings of the study suggest that fully transitioning SMEs into Industry 4.0 could lead to job creation, new business models, and internationalization of SME products. Moreover, regional cooperation and development in Africa may be essential to facilitate SMEs' involvement in 4IR technologies.

Vuong & Mansori (2021) conducted an empirical study to identify the impact of 4IR on Vietnamese enterprises, focusing on its effects on manufacturing, management, and operations. This study analyzed previous studies as secondary data and identified that 4IR adoption can create many challenges for Vietnamese enterprises such as impacting the way of manufacturing, management and operations. The study identifies those Vietnamese enterprises benefit from opportunities of 4IR like new technologies and a digital economy. However, they face challenges such as skill limitations, outdated technology, and intense competition. The study also suggested that enterprises should increase their investments and take innovative strategies to adopt and make proper use of 4IR technologies.

Antoniuk et al., (2017) conducted a study focusing on policy recommendations of Ukraine to improve the SME sector in the era of 4IR. For this purpose, the authors reviewed statistical reports of different institutions such as private companies, international organizations, NGOs and other existing studies on SMEs' innovative development. The findings of the study suggest that one of the major obstacles to SMEs' growth is the ineffective support provided by the government. To address this issue and offer guidance to Ukrainian policymakers, the authors conducted a survey among local high-tech SMEs. They identified key areas for improvement by using a SWOT analysis. Ultimately, a set of recommendations was developed to improve the SME environment in Ukraine, considering the challenges posed by 4IR.

In the context of Bangladesh, Friedrich-Ebert-Stiftung (FES) Bangladesh and SME Foundation Bangladesh jointly conducted a study focusing on ICT adoption in SME sector (Khan et al., 2021). The results of the study suggested that despite being a significant part of the economy and creating millions of jobs, SMEs in Bangladesh have been slow in adopting ICT. The study uses both primary and secondary research methods, including interviews and focus group discussions, to assess ICT adoption in various SME sectors. The manufacturing sector, particularly the plastic industry, shows more frequent usage of ICT tools compared to other industries in the sector. Some SMEs use very basic software like MS office (MS Word, MS Excel) and e-commerce sites for various business processes. On the other hand, the service sector has better ICT adoption, with logistics companies using vehicle tracking systems and the health industry using Enterprise Resource Planning (ERP) software. The identifies factors and challenges affecting ICT adoption, such as a lack of basic ICT knowledge among entrepreneurs, unskilled labor, and a lack of trust in local ICT service providers.

In another study (FES & SME, 2021), FES and SME Foundation Bangladesh collaboratively provided a policy brief on how to develop Technical and Vocational Education and Training (TVET) system to prepare ICT skilled manpower in Bangladesh. For this purpose, the German TVET system is studied as a role model to learn from this. The suggests that replicating the entire German model in Bangladesh may not be feasible due to differences in industrial maturity and limited financial resources in Bangladesh. Instead, the suggests a gradual adoption of key aspects from the German model which includes implementing apprenticeship programs, compensating trainees, providing free training in institutions, increasing government funding for educational institutions, and involving industry experts, chambers of commerce, and company representatives in curriculum design, examinations, and certification processes.

To the best of the researchers' knowledge, there is no study from these which directly investigates the impact of 4IR on the SME sector in Bangladesh. This kind of study is necessary, specifically for a developing country like Bangladesh, to identify and address the challenges and opportunities promised by 4IR.

## 3    Background

Revolution means abrupt and drastic change (Klaus Schwab, 2017). The term '*industrial revolution*' refers to a period of profound change in economic systems and social structures caused by the introduction of new technologies and novel ways of perceiving the world. These radical changes are divided into 4 categories namely the first, second, third and fourth industrial revolutions, as shown in Figure 1. The first industrial revolution was marked by the growth of railroads and the invention of the steam engine. The second industrial revolution was driven by the development of electricity and the assembly line. The third industrial revolution was triggered by the development of semiconductors, personal computers, and the Internet. These technologies also paved the way for the fourth industrial revolution, which is distinguished by much broader, deeper, and more complex and systemic applications of the preceding inventions, as well as the emergence of new technologies such as artificial intelligence and the Internet of Things (Rymarczyk, 2020).

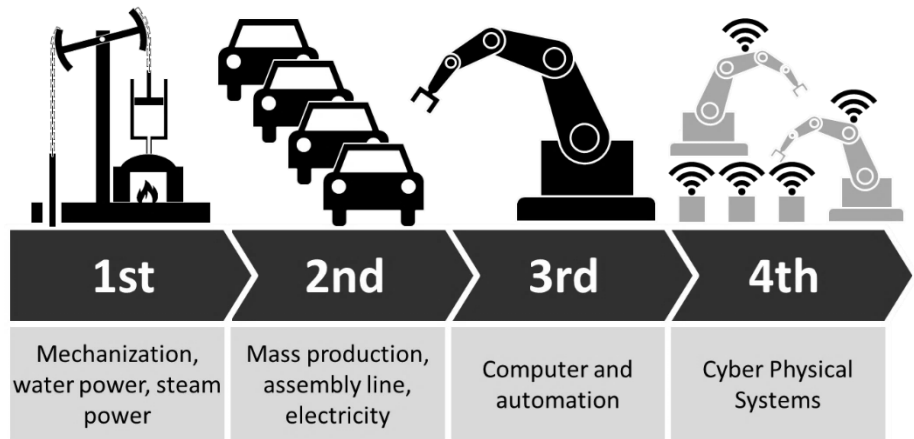

*Figure 1:The four industrial revolutions (by Christoph Roser at AllAboutLean.com)*

## 3.1 First Industrial Revolution

It began in the late 18th century, around 1760 to around 1840 (Klaus Schwab, 2017). Several factors including the development of new machinery, advancements in agriculture, and the growth of trade and commerce led to the start of the First Industrial Revolution. The steam engine was a notable innovation which powered industries and transportation. This revolution resulted in the transformation of an agrarian and handicraft economy into an industrial economy (Xu et al., 2018), leading to urbanization. In 1500, only 7% of the English population resided in cities, whereas it became 50% in 1850 (Allen, 2017). Although it brought about significant advancements and economic growth, it also brought challenges like social inequality and labor exploitation, particularly women and children receiving much lower wages than men (Mohajan, 2019).

## 3.2 Second Industrial Revolution

The Second Industrial Revolution, also known as the technological revolution, occurring from the mid-1800s to the early 1900s, marked a significant technological advancement with the introduction of electricity, the telephone, and the internal combustion engine (Afolalu et al., 2021). These innovations revolutionized communication, transportation, and daily life, while manufacturing techniques like the assembly line facilitated mass production of goods, leading to improved accessibility and affordability (Philbeck & Davis, 2019). Despite technological progress, the era was also characterized by harsh working conditions like long hours, low earnings (Mohajan, 2020) and dangerous environments in factories and mines, and increased unemployment due to the replacement of jobs by machines (Afolalu et al., 2021).

## 3.3 Third Industrial Revolution

The Third Industrial Revolution, often referred to as the digital revolution, began in the second part of the 20th century and continues to the present day. This revolution is characterized by the emergence of digital technologies such as the widespread use of computers, the birth of the internet, and the advent of automation. The rapid expansion of the Internet has facilitated unprecedented levels of worldwide communication and information exchange with over 4.95 billion people as of 2022 (Kurniasari et al., 2023). The advent of e-commerce has transformed the way goods and services are bought and sold and thus led to new economic models and opportunities.

Although these technologies increase productivity and efficiency, they also raise concerns about possible loss of jobs due to automation specially in sectors such as garments and furniture. Another important problem is the digital divide brought on by unequal access to technology and the internet, exacerbating disparities in educational and employment opportunities (Skilton & Hovsepian, 2017). Additionally, the rapid pace of technological change has posed regulatory and ethical challenges about the misuse of personal data and concerns about privacy

## 3.4 Fourth Industrial Revolution

The concept of the Fourth Industrial Revolution (4IR) was introduced in 2016 by Klaus Schwab, the founder of the World Economic Forum (Klaus Schwab, 2017). According to Zvika Krieger, the World Economic Forum's head of technology policy and partnerships, the Fourth Industrial Revolution differs from the Third for two reasons: (i) the distance between the digital, physical, and biological worlds is decreasing and (ii) technology is evolving faster than ever (CNBC, 2019). For example, it took 75 years for 100 million people to get access to the telephone, whereas the gaming app '*Pokemon Go*' intrigued many users in less than one month in 2016.

4IR is characterized by the rapid integration of technologies such as artificial intelligence, advanced robotics, the Internet of things, 3D printing, quantum computing and other technologies (Klaus Schwab, 2017) which are discussed below:

- **Artificial Intelligence (AI) and Machine Learning (ML)**: AI refers to the simulation of human-like intelligence processes by computers and machines (Salehi & Burgueño, 2018). ML is a subset of AI that focuses on developing algorithms and techniques that enable computers to learn from and make predictions or decisions based on data. These technologies are used in predictive analytics, natural language processing, and autonomous systems and are revolutionizing industries ranging from healthcare diagnostics to financial trading (Ferreira et al., 2021; Goldenberg et al., 2019). For example, speech and text recognition are already being used for communicating with patients and taking clinical notes (Davenport & Kalakota, 2019).

- **Internet of Things (IoT)**: IoT refers to the network of interconnected devices that are embedded with sensors, software, and network connectivity, enabling them to collect and share data by communicating with each other. These devices range from everyday household items like smart thermostats and wearables to industrial machinery and infrastructure systems. It has numerous applications including intelligent transportation, smart health care, smart home, etc. (Miazi et al., 2016). In business, IoT devices monitor parameters like temperature, humidity, air quality, energy, and machine performance, enabling real-time analysis to optimize operations and improve decision making process. Thus, IoT makes our world smarter and more connected by converting ordinary objects to smart objects with the ability to communicate and collaborate.

- **Advanced robotics and collaborative robots**: Advanced robots enable humans to avoid monotonous, risky or challenging tasks. Robots are now being deployed in industries like manufacturing, healthcare, and agriculture. For instance, some hospitals are already using robot assistants (Oosthuizen, 2022). Collaborative robots, also known as cobots, are being used to assist humans in dangerous or labor-intensive tasks such as lifting or delivering large or hazardous commodities while being supervised by humans (Fanoro et al., 2021) .

- **3D printing**: It is a process that uses computer-aided design data to create a physical object through the method of additive manufacturing (Beitler et al., 2022). 3D printing has a wide range of applications including the automotive manufacturing industry, the biomedical industry, the healthcare sector and the construction industry. It is being used in healthcare products such as hearing aids, artificial ears, rehabilitation aids, artificial joints and personalized dental implants (Wang et al., 2019).

- **Virtual and Augmented Reality (VR/AR)**: VR creates a fully immersive digital environment that simulates the physical presence of the user in a computer-generated world (Gandhi & Patel, 2018) . AR enhances the real-world environment by overlaying virtual computer-generated elements onto it (Carmigniani & Furht, 2011). VR and AR have already been applied to various fields, for example, military, manufacturing, medical, gaming, advertising and entertainment (Nee & Ong, 2013).

- **Quantum Computing**: It is a cutting-edge technology that harnesses the principles of quantum mechanics to perform computations in a fundamentally different way from traditional computers (Boughzala et al., 2022). It has the potential to revolutionize fields such as cryptography, optimization, material science, drug discovery, etc. For instance, global shipping companies are using quantum computers to optimize shipping routes, reducing fuel consumption and environmental impact (Right People Group, n.d.).

- **Biotechnology and Genetic Engineering**: Modern biotechnology developed with genetic engineering and the sequencing of the human genome in 2001 (Andreoni et al., 2021). In recent years, significant progress has been achieved in lowering the cost and increasing the ease of genetic sequencing as well as more in activating or editing genes. A genome may now be sequenced in a matter of hours and for less than a thousand dollars (Klaus Schwab, 2017). With increases in processing power, scientists no longer rely on trial and error; instead, they investigate how specific genetic mutations produce certain features and diseases.

## 3.5   Impact of 4IR on Employment

The Fourth Industrial Revolution (4IR) has brought about significant changes in the employment landscape, particularly in the context of SMEs. While 4IR technologies, such as automation, artificial intelligence, and the Internet of Things, have the potential to streamline operations and enhance productivity in SMEs, they also pose challenges to traditional employment patterns.

The current and the next decade are expected to face significant challenges for employment across the world. The World Economic Forum (a2i, 2023) forecasts that half of all employees worldwide will require reskilling by 2025. Along with many other countries, Bangladesh will also encounter major challenges for future employment as a result of 4IR with automation.

According to the World Economic Forum (2018), 75 million jobs will be displaced. But at the same time, the report also estimates that a possible 133 million new jobs will be created. A study by a2i project (a2i, 2020) suggests that 47% of jobs in Bangladesh could be at risk by 2041. These also includes several sectors of SME. Based on the collaborative research effort between the Government's a2i initiative and the International Labour Organization (ILO) (Elder & Emmons, 2018), it is projected that by 2041, the Tourism and hospitality sector will witness a decline of 20% (6 lakh), the Leather Industry will experience a reduction of 35% (1 lakh), Agricultural Products will see a decrease of 40% (6 lakh), the Furniture Sector will encounter a decline of 55% (13.8 lakh), and the Garments Sector will be affected the most with a decline of 60% (27 lakh) due to automation. However, on the bright side, while the automation of 4IR could lead to the loss of 5.5 million job, the same technological revolution also holds the potential of generating 10 million new jobs (a2i, 2023) .

The Center for Policy Dialogue (CPD) (CPD, 2018) conducted a study in the Readymade Garment sector, revealing a 16% reduction in the number of female workers at the officer level. Female workers often lack the same level of knowledge about operating machines as their male counterparts, which contributes to fewer women in managerial positions. Additionally, social, psychological, and cultural obstacles limit women's participation in skill development activities and contribute to such skill gap hindering their promotion opportunities. Women also face difficulties in advancing to higher-paid, secure jobs due to domestic duties and inflexible work conditions.

The period between 2005 and 2012 saw a 4.01% increase in employment growth, while from 2012 to 2016, this growth rate decreased to 3.3%. According to the Policy Research Institute of Bangladesh (PRI) (The Financial Express, n.d.), there were 545 employed workers in 1990 for every million dollars in export, but in 2016 this dwindled to 142 because of automation. It can worsen income inequality in society, as high-skill, high-wage jobs grow while low-skill jobs decline. The inequality can increase among people, as not all sectors or workers may equally benefit from technological advancements. Moreover, regions with strong technological infrastructure and innovation ecosystems may thrive, while others lag, leading to geographic inequality. Continuous learning, upskilling, and adaptability will be crucial factors for employment as technological advancements rapidly change job requirements.

The 4IR is closely linked to the changes in employment discussed above. While automation and technological advancements due to 4IR can reduce the need for certain jobs, they also create new opportunities in emerging industries and roles. The overall impact on employment varies across different regions, sectors, and skill levels and is influenced by factors such as globalization, demographic changes, and government policy. Balancing potential job losses with job creation and ensuring that workers have the necessary skills to transition into new roles are crucial challenges in the Fourth Industrial Revolution era.

In Bangladesh, approximately 30 percent of the total employed workforce is engaged SMEs across various sectors, including Ready-Made Garments (RMG), Agriculture, Leather Industry, Furniture etc. Therefore, the impact of the 4IR on employment, as indicated by the statistics discussed earlier, is likely to be most significant in the SME sector. To summarize, 4IR can lead to the displacement of certain low-skilled jobs within SMEs that can be replaced with automation, which may result in short-term job losses. However, 4IR will also create opportunities for SMEs to adapt new roles in tech-related fields by upskilling their workforce. Additionally, the digital transformation brought about by 4IR can open doors to global markets, potentially leading to job growth in areas such as e-commerce, digital marketing, and online customer support. Therefore, the impact of 4IR on employment in the SME sector is a complex interplay of challenges and opportunities, ultimately depending on the ability of SMEs to adapt and harness the full potential of these technologies.

## 4  Methodology

This paper aims to analyze the impact of 4IR on SMEs and employment in Bangladesh. For this purpose, a qualitative study has been performed. It denotes the paper of the nature of phenomena in a non-numerical format such as text, audio or video (Busetto et al., 2020; Gelo et al., 2008). It helps to gather an in-depth understanding of a problem or generate new ideas for further research. This study used a combination of primary and secondary data sources to ensure a thorough investigation. Primary data were collected from both Key Informant Interviews and Focus Group Discussions to overcome the weaknesses of each single method (Sife et al., 2010). Secondary data were collected from published reports, journals and websites.

The collected data were examined using thematic analysis (Le et al., 2020). The overview of the methodology is shown in Figure 2.

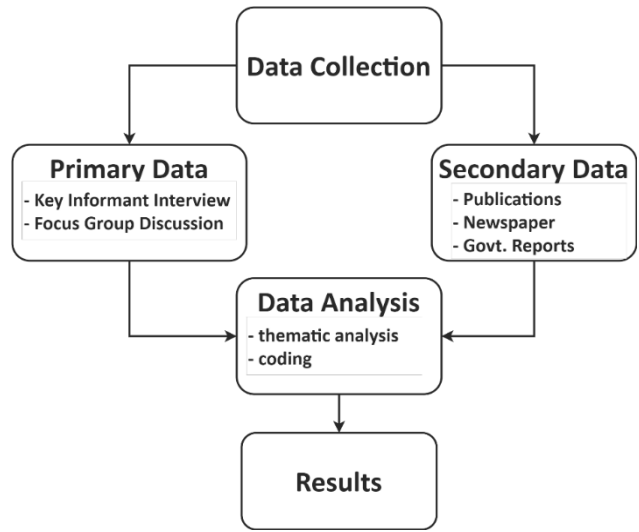

Figure 2:Overview of the Methodology

## 4.1 Secondary Data Collection

Newspaper and magazine articles, journal and conference articles, government reports as well as reports from other national and international organizations were reviewed to identify the existing benefits, challenges of 4IR and policies related to it. Next, a questionnaire was prepared by incorporating these factors.

## 4.2 Primary Data Collection

For collecting primary data, both Key Informant Interviews and Focus Group Discussions were performed to complement each method (Adelowotan, 2021). In this phase, the following steps were executed.

I. **Designing sample**: To collect data, at first, major stakeholders of the SME industry were identified using prior literature. According to (Clarke & Braun, 2014), SME owners, association representatives and relevant representatives from the national policymakers and government bodies were found to be the primary stakeholders.

SME Foundation provided the researchers with a list of SMEs which were grouped based on the eleven booster sectors identified by the Ministry of Industries, Government of Bangladesh (Ahmed & Chowdhury, 2009).

1) Electronics and Electrical
2) Software-development
3) Light engineering and metal-working
4) Agro-processing/agro-business/plantation agriculture/specialist farming/tissue-culture
5) Leather-making and leather goods
6) Knitwear and ready-made garments
7) Plastics and other synthetics
8) Healthcare & diagnostics
9) Educational services
10) Pharmaceuticals/cosmetics/toiletries

11) Fashion-rich personal effects, wear and consumption goods

Next, the participants were randomly chosen from this list.

**II. Conducting Key Informant Interview**: Based on the questionnaire prepared in 4.1, Key Informant Interviews were performed. Most of the interviews were conducted over the phone and recorded. The researchers also took notes during the interviews, which were later cross-checked with the recordings for accuracy. Based on the responses of the participants, an updated questionnaire was prepared which was used for Focus Group Discussion. The questionnaire can be found in Appendix – 1: Questionnaire.

**III. Conducting Focus Group Discussion**: The Focus Group Discussion involved top management staffs from various SME sectors who were also randomly chosen from the SME Foundation list. The session was conducted by the primary researcher and lasted around two and a half hours.

## 4.3  Data Analysis

After collecting data from primary and secondary sources, those were analyzed using thematic analysis (Adegbite & Govender, 2021). Initially, two researchers independently coded the transcripts of the Key Informant Interview and Focus Group Discussion (Ohta et al., 2020). They then discussed their results until they reaching a consensus. Based on the analysis, opportunities and challenges of 4IR for SMEs were identified. and corresponding policy recommendations were made.

## 5  Result Analysis

In this study, 16 people individuals participated in the KII and FGD. Among them, 12 were SME owners, two were association representatives and two were representatives from the national policymakers and government bodies, as shown in Figure 3. The detailed information about the participants can be found in Appendix – 2: Participants List.

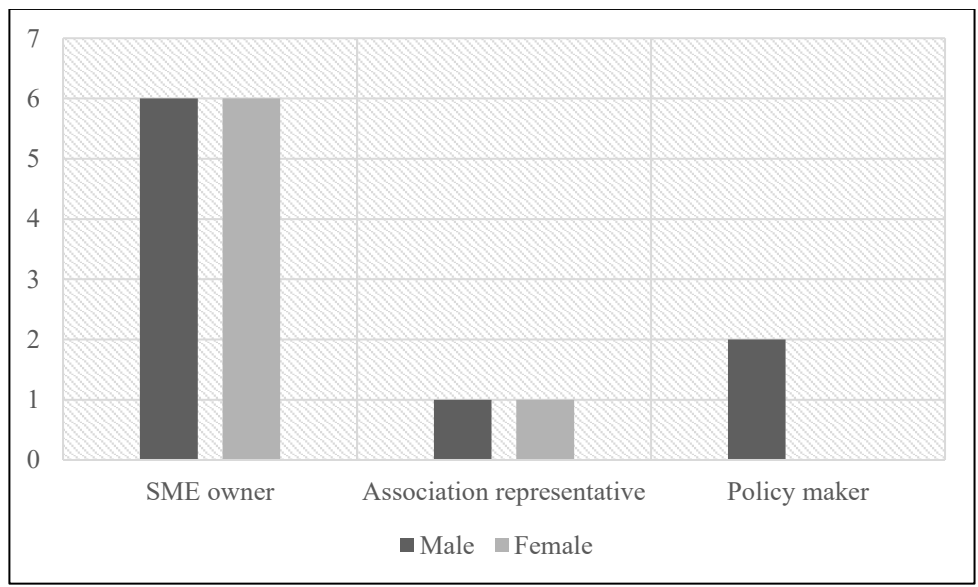

*Figure 3: Details of the Participants*

The study found that most of the respondents have a surface-level awareness of 4IR without a detailed understanding. Despite this, they consider 4IR as a blessing. Although they have not yet fully embraced its potential, they are hopeful that it will provide them with several benefits:

- **Enhancing customer experience**: 4IR will better meet customer requirements and facilitate mass customization of products. It will enable business to offer a more personalized, efficient, and quality-driven customer experience through advanced technologies. Reducing production time and increasing profitability: 4IR related tools and technologies will automate workflows, minimizing manual interventions. Thus, it will allow business entities to use their resources more effectively to reduce production time and eventually lead to increased profitability.

- **Improving product quality**: 4IR is anticipated to improve product quality through the use of modern machinery and advanced tools for real-time product monitoring. These improvements will minimize defects and errors resulting high quality of the products.

- **Data-driven insights**: 4IR will help SMEs collect and analyze data regarding market trends, customer preferences and operational performance. These data-driven insights will enable SMEs to make informed decision, leading to strategic business improvements.

In the employment sector, participants believe that 4IR will bring significant changes that can be summarized two main aspects:

- **Creating new job opportunities and business**: 4IR is characterized by the rapid advancement and integration of cutting-edge technologies such as Artificial Intelligence (AI), the Internet of Things (IoT), blockchain, 3D printing. These technologies enable the creation of innovative products, services, and business models that were previously unimaginable. According to a study by a2i (2022), new occupations are expected to emerge in the prominent sectors of SME such as RMG and Textile, Furniture, Agro-food Processing, Leather and Tourism. For example, sectors like RMG and Textile manufacturing and furniture in Bangladesh, where low-skilled labor is predominant, are expected to face significant job losses due to automation. This will create new job opportunities with a greater demand for skilled and semi-skilled workers. Bangladesh can win in these sectors if adequate measures are taken to upgrade skills and embrace new technologies. This finding is aligned with the World Economic Forum's predictions that suggests around 133 million new jobs will be created due to technological advancement (Oosthuizen, 2019).

- **Increasing the productivity among workers**: 4IR has the potential to significantly enhance worker productivity through the adoption of advanced technologies, automation, and the optimization of work processes. 4IR technologies, such as robotics and advanced automation systems can take over repetitive tasks, enabling employees to focus on more value-added activities that require creativity, critical thinking and thus increasing overall productivity.

Respondents express the belief that employees will not lose their jobs due to the 4IR; rather, their roles may undergo transformation. Owners mostly predominantly receive positive feedback from the employees regarding 4IR, noting their interest in learning new skills to enhance their salary or other benefits. While, adaption may require time among workers, owners acknowledge their responsibility to prepare their workers.

Despite the overall positive statement towards 4IR, only one of the SME owners currently utilizes 4IR -related tools and technologies such as ChatGPT, Copilot - in their organization. This disparity is attributed to several challenges:

- **Lack of knowledge**: Many SME owners do not have knowledge of existing 4IR tools and technologies and how to integrate them into their company/businesses. For example, one of the SME owners was having trouble marketing her products due to unawareness of available tools. Besides, some SME owners have a misconception that such 4IR-related tools and technologies are only necessary for large organizations.

- **Financial issues**: 4IR-related tools and technologies often are expensive, posing financial challenges for SMEs. While, one of the policymakers mention the availability of loans, SMEs owners express frustration over the difficulty in accessing them.

- **Lack of proper training**: Some of the SME owners cite a lack of training opportunities or workshop related to 4IR as a barrier. Some other pointed out that though the government is organizing skill development training, but those are perceived to be inadequate due to issues such as insufficient knowledge of the trainers, and budget constraints.

- **Safety and security related issues**: SME owners also have concerned about the safety and security implications of adopting 4IR technologies. For example, one of the SME owners fears of plagiarism which hindering adoption of modern technology.

- **Infrastructure limitations**: Many participants highlight insufficient infrastructure, particularly concerning internet availability and speed, as a barrier to utilizing 4IR technologies effectively.

- **Information accessibility**: A lack of a dedicated platform for SMEs for exchange information, hindering their ability to access assistance or guidance. Employees, despite the interest, they do not know whom to contact in case of any help. Furthermore, although workers are interested in learning modern tools and technologies, struggle to find suitable resources.

Several steps have already been initiated to address these challenges faced by the SEMs such as the establishment of one-stop digital service centers (The Daily Star, n.d.) . However, respondents suggest additional measures for further improvement. The following steps can be taken in adopting to the 4IR:

- **Thorough analysis of current situation**: Before moving towards 4IR adoption, it is essential to comprehensively analysis the current situation of SMEs. The analysis should identify sectors that require special attention.

- **Creation of a platform for SMEs**: A platform can be created where SMEs can share information, report their problems, get suggestions can be beneficial. This platform could facilitate finding suitable markets for their products, and obtaining guidance how to promote their products or promotion strategies.

- **Organizing training and workshops**: To prepare the SMEs for utilizing 4IR, comprehensive training and workshop are crucial comprehensive. These sessions should provide both theoretical and hands-on experience with 4IR related tools and technologies. Therefore, quality training and workshops need to be organized and to verify its implications, there should be arrangements for follow-ups.

- **Provision for sharing tools**: Since modern types of machinery and tools can be expensive at times. The platform created for SMES could facilitate collaborative arrangements for tool sharing such as similar types of SMEs can be grouped together would enable them to share resources, tools and reduce their cost.

- **Attracting foreign clients**: The government and industry associations should undertake steps to help SMEs attract foreign clients, in addition to the local clients, thereby expanding market technologies.

- **Development of indigenous tools and technologies**: To reduce dependence on imported machinery and maintaining lower costs, Bangladesh should invest in creating its own tools and technologies.
- **Raising nationwide awareness**: Since 4IR will have an impact on everyone - SME owners, workers, clients and associations, a nationwide awareness campaign need. This campaign should highlight benefits of 4IR adoption and provide strategies for overcoming associated challenges.

Based on the results from this study it is evident that a further detailed study and large-scale data collection is needed. As participants expressed their concern of adequate training and workshops on 4IR, indicating that capacity building in this sector is necessary to support them. Many participants expressed their inability to adopt ICT in SME sector without the cooperation from policy level, a shared resource platform can play an important role in this case. The national policy needs to provide this support in terms of finance or resource need.

Apart from creating policies, the government should ensure they are effectively implemented and also monitor that they are being followed. Besides, there has to be the continuation of policy even when there are changes in administration.

Finally, the government, associations and academicians must collaborate to ensure that 4IR becomes a beneficial for SMEs.

# 6 Conclusion

The study investigates the opportunities and challenges presented by the Fourth Industrial Revolution (4IR) for small and medium-sized enterprises (SMEs). Through the interviews and discussions with SME owners, association representatives, and policymakers, the research reveals that while most of the respondents have basic of 4IR, they lack a detailed understanding of its implications of SMEs. The potential benefits of 4IR include enhancing customer experience, reducing production time, improving product quality, and enabling data-driven decision-making. Additionally, the study indicates that 4IR holds the promise of creating new job opportunities and boosting worker productivity.

Despite these advantages, the study identifies several barriers that could prevent SMEs from fully embracing and utilizing 4IR technologies. These challenges include a lack of knowledge, financial constraints, inadequate training, safety and security concerns, inadequate infrastructure, and a lack of information-sharing platforms. To overcome these challenges, the study recommends various measures such as analyzing the current situation of SMEs, creating a platform for information sharing, organizing training sessions and workshops, and fostering domestic technology development.

Finally, the successful integration of 4IR technologies into SME operations necessitates a collaborative effort involving government bodies, industry associations, academia and Civil Society Organizations (CSOs). It also underscores the importance of raising nationwide awareness about the implications of 4IR, ensuring effective policy implementation and maintaining continuous policy monitoring and adjustment. In future, large-scale sectoral data can be collected to understand the challenges and opportunities across various sub-sectors of SMEs in adopting 4IR.

# Acknowledgement

We gratefully acknowledge the support of the SME Foundation Bangladesh and the Friedrich-Ebert-Stiftung (FES) Bangladesh. The financial support and the constructive feedback from both organizations have made the study complete and successful.

# Appendix – 1: Questionnaire

| Name (নাম) : | |
|---|---|
| Address (ঠিকানা) : | |
| Designation (পদবি) : | |
| Gender (লিঙ্গ) : | |
| Level of education: (শিক্ষাগত যোগ্যতা) | • Primary (প্রাইমারি)<br>• Secondary (সেকেন্ডারি)<br>• Graduation (গ্র্যাজুয়েশন)<br>• Post Graduation (পোস্ট গ্র্যাজুয়েশন)<br>• PhD (পিএইচডি)<br>• Other (অন্যান্য) _______ |
| Email (ই-মেইল) : | |
| Phone (ফোন): | |

| |
|---|
| Are you aware of the 4th Industrial Revolution (4IR)? (আপনি ৪র্থ শিল্প বিপ্লব (4IR) সম্পর্কে অবগত/সচেতন কিনা?) |
| What kind of tools & technology related to 4IR (software/tools/machines) are you using in your organization? (আপনি আপনার প্রতিষ্ঠানে 4IR সম্পর্কিত কোন ধরনের প্রযুক্তি (সফ্টওয়্যার /টুল/ মেশিন) ব্যবহার করছেন?) |
| Is it a blessing or a threat to SMEs? (৪র্থ শিল্প বিপ্লব (4IR) কি এসএমই খাতের জন্য আশীর্বাদ নাকি হুমকিস্বরূপ?) |
| What benefits is 4IR providing to SMEs? (৪র্থ শিল্প বিপ্লব (4IR) এসএমই খাতে কী কী সুবিধা প্রদান করছে?)<br>_Hints (সংকেত)_:<br>_Improving customer experience (গ্রাহক অভিজ্ঞতা উন্নত করা)_<br>_Reducing production time (উৎপাদন সময় কমানো)_<br>_Increasing profitability (লাভজনকতা বৃদ্ধি)_<br>_Others (অন্যান্য)_ |
| To be specific, what benefits is 4IR providing particularly in the employment sector? (বিশেষত এসএমই-এর কর্মসংস্থান খাতে ৪র্থ শিল্প বিপ্লব (4IR) কী কী সুবিধা প্রদান করছে?)<br>_Hints (সংকেত)_:<br>_Creating new job opportunities (নতুন কাজের সুযোগ সৃষ্টি করা)_<br>_Retaining existing employees (বিদ্যমান কর্মচারীদের ধরে রাখা)_<br>_Others (অন্যান্য)_ |
| What challenges are 4IR introducing to SMEs? (৪র্থ শিল্প বিপ্লব (4IR) এর কারণে এসএমই খাত কী কী চ্যালেঞ্জের মুখোমুখি হচ্ছে?)<br>_Hints (সংকেত)_:<br>_Safety and security related issues (নিরাপত্তাজনিত সমস্যা)_<br>_Financial issues (আর্থিক সমস্যা)_<br>_Infrastructure related issue (অবকাঠামোগত সমস্যা)_<br>_Management related issues (ব্যবস্থাপনা সংক্রান্ত সমস্যা)_ |
| Lack of innovations through research and development (R & D) (গবেষণা ও উন্নয়নের মাধ্যমে উদ্ভাবনের অভাব) <br> *Increasing dominance of MNC (Multinational Companies) over SME (এসএমই-এর উপর মাল্টিন্যাশনাল কোম্পানির ক্রমবর্ধমান আধিপত্য)* <br> *Others (অন্যান্য)* |
| To be specific, what challenges is 4IR introducing in the employment sector? (বিশেষত এসএমই-এর কর্মসংস্থান খাতে 4IR কী কী চ্যালেঞ্জ প্রবর্তন করছে?) <br> <u>Hints (সংকেত)</u>: <br> *Shortage of skilled knowledge workers (দক্ষ কর্মীর অভাব)* <br> *Resistance from workers in adopting new technologies (নতুন প্রযুক্তি গ্রহণে শ্রমিকদের থেকে বাধা)* <br> *Fear of losing jobs among workers (শ্রমিকদের মধ্যে কাজ হারানোর ভীতি)* <br> *Others (অন্যান্য)* |
| What policies by the Government/associations/organizations have already been made so far or can be made in the future to help SMEs to cope with 4IR? (৪র্থ শিল্প বিপ্লব (4IR) -এর চ্যালেঞ্জ মোকাবেলায় সরকার/এসএমই অ্যাসোসিয়েশনসমূহ/ প্রতিষ্ঠানসমূহ এখন পর্যন্ত কী কী নীতিমালা প্রনয়ণ করেছে এবং ভবিষ্যতে আরও কী কী করা যেতে পারে?) <br> <u>Hints (সংকেত)</u>: <br> *Access to Finance (আর্থিক সহজ লভ্যতা)* <br> *Access to Technology and Innovation (প্রযুক্তি এবং উদ্ভাবনে সহজ লভ্যতা)* <br> *Access to Market (মার্কেটের সহজ লভ্যতা)* <br> *Access to Education and Training (শিক্ষা ও প্রশিক্ষণের সহজ লভ্যতা)* <br> *Access to Business Support Services (ব্যবসায়িক সহায়তা পরিষেবার সহজ লভ্যতা)* <br> *Access to Information (তথ্যের সহজ লভ্যতা)* <br> *Others (অন্যান্য)* |
| Anything else that you want to share? (আপনি কি এ সম্পর্কে আরও কিছু যোগ করতে চান?) |

## Appendix – 2: Participants List

The participant list is available upon request, subject to ethical and privacy guidelines.